\documentclass[12pt]{iopart}
\usepackage{iopams}
\usepackage{latexsym}
\usepackage{color}

\newcommand{\ad}{\mathrm{ad}}

\newcommand{\eqref}[1]{(\ref{#1})}

\begin{document}

\title[On rational R-matrices with adjoint SU(n) symmetry]{On rational R-matrices with adjoint SU(\emph{n}) symmetry}

\author{Laurens Stronks,$^{1,2}$ Johan van de Leur$^2$ and Dirk Schuricht$^1$}

\address{$^1$Institute for Theoretical Physics, Center for Extreme Matter and Emergent Phenomena, Utrecht University, Princetonplein 5, 3584 CE Utrecht, The Netherlands}
\address{$^2$Mathematical Institute, Utrecht University, Budapestlaan 6, 3584 CD Utrecht, The Netherlands}

\ead{d.schuricht@uu.nl}

\begin{abstract}
Using the representation theory of Yangians we construct the rational R-matrix which takes values in the adjoint representation of SU(\emph{n}). From this we derive an integrable SU(\emph{n}) spin chain with lattice spins transforming under the adjoint representation. However, the resulting Hamiltonian is found to be non-Hermitian. 
\end{abstract}

\section{Introduction}\label{sec:introduction}
Consider a finite-dimensional, irreducible representation $\rho: g\to\mathrm{end}(V)$ of some complex Lie algebra $g$ on the vector space $V$. An R-matrix $R(\lambda): V\otimes V\to V\otimes V$ for $\lambda\in\mathbb{C}$ is a solution of the Yang--Baxter equation 
\begin{equation}
R_{12}(\lambda)\,R_{13}(\lambda+\mu)\,R_{23}(\mu)=R_{23}(\mu)\,R_{13}(\lambda+\mu)\,R_{12}(\lambda)
\label{eq:YBE}
\end{equation}
on the tensor product $V\otimes V\otimes V$, where $R_{ij}(\lambda)$ acts non-trivially on the $i$-th and $j$-th factor, eg, $R_{12}(\lambda)=R(\lambda)\otimes 1$. Solutions of the Yang--Baxter equation play a central role in various fields of theoretical physics like the quantum inverse scattering method~\cite{KorepinBogoliubovIzergin93}, statistical field theories with factorised scattering~\cite{Mussardo10,SamajBajnok13}, the theory of quantum groups~\cite{Ma93,ChariPressley98}, or quantum information theory~\cite{Dye04,KauffmanLomonaco04}.

In the early 1980s many rational R-matrices beyond the simplest case of the fundamental representation of su(2) have been constructed, see, eg, Refs.~\cite{Kulish-81,KulishSklyanin82,Reshetikhin85,KulishReshetikhin86,AlishauskasKulish86}. In 1985, Drinfel'd~\cite{Drinfeld85,Drinfeld88} then discovered the intimate relation between rational R-matrices and Yangians Y($g$). He proved that there exists a universal R-matrix in $\mathrm{Y(}g\mathrm{)}\otimes\mathrm{Y(}g\mathrm{)}$ which yields rational solutions of \eqref{eq:YBE} provided the representation $\rho$ of $g$ can be extended to an irreducible representation of Y($g$). Unfortunately the universal R-matrix is not known explicitly. This limitation was overcome by Chari and Pressley~\cite{ ChariPressley90,ChariPressley91}, who gave an explicit construction of rational R-matrices based on the intertwining operator of representations of the Yangian. We will apply this method to construct rational solutions of \eqref{eq:YBE} where $V$ is the $(n^2-1)$-dimensional, adjoint representation of su($n$), and discuss the physics implication of our result. While this is presumably well-known to the integrable-systems community, it does not seem to have been published before.

This article is organised as follows: In the next section we review the necessary facts about Yangians and their relation to R-matrices. In Sec.~\ref{sec:Adjrep} we present basic results on the adjoint representation of su(\emph{n}) and the properties of its tensor products. We then use these results to construct the rational R-matrix for the adjoint representation of su(\emph{n}), which constitutes our main result and is given in \eqref{eq:rmat}. In Sec.~\ref{sec:hamiltonian} we construct the corresponding integrable spin-chain Hamiltonian, which, however, turns out to be non-Hermitian. Finally, in Sec.~\ref{sec:SU3} we treat the special case of SU(3), before we conclude with a discussion of our results.

\section{Yangians and R-matrices}\label{sec:Yangian}
We first review the relation between Yangians and R-matrices based on the intertwining operator~\cite{ ChariPressley90,ChariPressley91}. The Yangian Y(su(\emph{n})) is the infinite-dimensional Hopf algebra generated by the (level-0) elements $x\in\mathrm{su(}n\mathrm{)}$ and the level-1 elements $J(x)$, which satisfy a set of non-trivial relations that can be found, eg, in Ref.~\cite{ChariPressley91}. For example, the map $J: \mathrm{su(}n\mathrm{)}\to\mathrm{Y(su(}n\mathrm{))}$ is linear and satisfies $[x,J(y)]=J([x,y])$. In particular, the Yangian possesses a co-multiplication $\Delta: \mathrm{Y(su(}n\mathrm{))}\to\mathrm{Y(su(}n\mathrm{))}\otimes\mathrm{Y(su(}n\mathrm{))}$ which is given by
\begin{equation}
\fl\quad
\Delta(x)=x\otimes 1+1\otimes x,\quad\Delta(J(x))=J(x)\otimes 1+1\otimes J(x)+\frac{1}{2}\Bigl[x\otimes 1,\Omega\Bigr]
\label{eq:comultiplication}
\end{equation}
with the Casimir element $\Omega\in\mathrm{su(}n\mathrm{)}\otimes\mathrm{su(}n\mathrm{)}$. Choosing an orthonormal basis $I^a$, $a=1,\ldots, n^2-1$, of su(\emph{n}) (with respect to the trace form $(I^a,I^b)=\mathrm{tr}[(I^a)^\dagger I^b]=\delta^{ab}$), we find $\Omega = \sum_a I^a \otimes I^a$ and the level-1 generators are given by $J^a=J(I^a)$. For the orthonormal basis $\{I^a\}$ one can, up to a normalisation, use the generalised Gell-Mann matrices~\cite{BertlmannKrammer08}. We recall that in such a basis the commutator and anticommutator of two elements can be written as
\begin{equation}
\Bigl[I^a, I^b\Bigr] = \sum_cf^{abc} I^c,\quad \Bigl\{I^a, I^b\Bigr\} = \sum_cd^{abc} I^c + \frac{2}{n} \delta^{ab}\,1,
\label{eq:fabc}
\end{equation}
where $f^{abc}$ are the antisymmetric, purely imaginary structure constants, $d^{abc}$ the symmetric, real $d$-symbols, and 1 denotes the $(n\times n)$-identity matrix. 

Starting from a representation $\rho: \mathrm{su}(n)\to\mathrm{end}(V)$ of su(\emph{n}), one can construct representations of the corresponding Yangian using the evaluation homomorphism~\cite{ChariPressley98} ev$_\lambda: \mathrm{Y(su(\emph{n}))}\to \mathrm{U(su}(n))$ for $\lambda\in\mathbb{C}$, which is given for $n\ge 3$ by
\begin{equation}
\mathrm{ev}_\lambda(x)=x,\quad \mathrm{ev}_\lambda(J(x))=\lambda x+\frac{1}{4}\sum_{a,b}\mathrm{tr}\Bigl(x\{I^a,I^b\}\Bigr)I^a I^b.
\label{eq:evahom}
\end{equation}
Now the composition $\rho_\lambda=\rho\circ\mathrm{ev}_\lambda$ defines a one-parameter family of representations of the Yangian which we denote by $V_\lambda$ for simplicity. 

Taking two representations $V_\lambda$ and $W_\mu$ of Y(su(\emph{n})), the action on their tensor product can be defined using the co-multiplication \eqref{eq:comultiplication}. Chari and Pressley proved~\cite{ ChariPressley90,ChariPressley91}  that there exists a unique intertwining operator $I_{V,W}(\lambda-\mu): W_\mu \otimes V_\lambda \rightarrow V_\lambda \otimes W_\mu$ which is compatible with the action of the Yangian and preserves the tensor product of the highest-weight states in $V_\lambda$ and $W_\mu$. Furthermore, the intertwining operator is a rational function of $\lambda-\mu$ and satisfies $I_{V,W}(\lambda-\mu)\,I_{W,V}(\mu-\lambda)=1$. Most importantly, they showed that the intertwining operator yields a rational R-matrix, ie, $R(\lambda)=I_{V,V}(\lambda)\sigma$ with the permutation operator $\sigma$ satisfies the Yang--Baxter equation. In Sec.~\ref{sec:Rmat} we will apply this construction to obtain the R-matrix for the adjoint representation of su(\emph{n}). 

\section{Adjoint representation of su(\emph{n})}\label{sec:Adjrep}
Before doing so we recall some necessary facts about the adjoint representations. For convenience, we assume $n>3$ and treat the case $n = 3$ separately in Sec.~\ref{sec:SU3}.

The Lie algebra su(\emph{n}) possesses $n-1$ simple roots $\alpha_i$, $i=1,\ldots, n-1$, in terms of which the highest weight of the adjoint representation is given by the maximal root $\beta = \sum_{i=1}^{n-1}\alpha_i$. Alternatively this can be written as~\cite{Humphreys72,MacfarlanePfeiffer00} $\beta = \lambda_1 + \lambda_{n-1}$, there the $\lambda_i$ are the fundamental weights. In general, any representation can be uniquely defined by its highest weight $\sum_{i=1}^{n-1} m_i \lambda_i$ with integers $m_i$; we will use the notation $(m_1 m_2 \cdots m_{n-1})$ to specify the corresponding representation. 

The tensor product representation of two adjoint representations decomposes as
\begin{eqnarray}
&&(1 0 \ldots 0 1) \otimes (1 0\ldots 0 1)= (2 0 \ldots 0 2)_\mathrm{s} \oplus (2 0 \ldots 0 1 0)_\mathrm{a} \oplus (0 1 0  \ldots  0 2)_\mathrm{a}  \nonumber
 \\ &&\qquad\qquad\quad \oplus  (0 1 0  \ldots 0 1 0)_\mathrm{s} \oplus (1 0 \ldots 0  1)_\mathrm{s} \oplus (1 0 \ldots  0  1)_\mathrm{a} \oplus (0 \ldots  0)_\mathrm{s},\label{eq:sundec}
\end{eqnarray}
where $(010\ldots 010)$ has to read $(020)$ for $n=4$. With the subindices we indicate the symmetry properties under exchange of the factors in the tensor product. 

The positive root spaces of su(\emph{n}) can be chosen such that they correspond to the one-dimensional subspaces spanned by $e_{ij}$ with $i < j$, where $e_{ij}$ is the matrix with $1$ as the $(i,j)$-th entry and $0$ otherwise. With this choice of positive roots, we fix the root space $g_{\alpha_1}$, with $\alpha_1$ a root corresponding to one of the end nodes in the Dynkin diagram, to be spanned by $e_{12}$. The highest-weight vector of the adjoint representation is then given by $e_{1n}$. For later use we state the explicit expressions for the highest-weight vectors of the submodules on the right-hand side of \eqref{eq:sundec} in Table~\ref{tab:hwv2}, which can easily be checked by straightforward calculation. For example, from \eqref{eq:fabc} one sees that $\{\Omega, 1 \otimes e_{1n}\} -\frac{2}{n} e_{1n} \otimes 1$ indeed belongs to $\mathrm{su(\emph{n})}\otimes\mathrm{su(\emph{n})}$.
\begin{table}[t]  
\centering 
\begin{tabular}{|l||l||l|} 
		\hline Submodule & Highest-weight vector & Eigenvalue of $\Omega$ \\
		\hline $(2 0 \cdots 0 2)_\mathrm{s}$ & $e_{1n} \otimes e_{1n}$ & $2$ \\ 
		\hline $(2 0 \cdots 0 1 0)_\mathrm{a}$ & $e_{1(n-1)} \wedge e_{1n}$ & $0$\\
		\hline $(0 1 0 \cdots 0 2)_\mathrm{a}$ & $e_{2n}\wedge e_{1n}$& $0$ \\
		\hline $(0 1 0  \cdots 0 1 0)_\mathrm{s}$ & $\mathrm{Sym} \left(e_{2(n-1)} \otimes e_{1n} - e_{1(n-1)} \otimes e_{2n} \right)$& $-2$ \\
		\hline $(1 0 \cdots  0 1)_\mathrm{s}$ & $ n \left(\{\Omega, 1 \otimes e_{1n}\} -\frac{2}{n} e_{1n} \otimes 1 \right)$& $-n$ \\
		\hline $(1 0 \cdots 0 1)_\mathrm{a}$ & $\sqrt{n^2-4}\,[\Omega,1 \otimes e_{1n}]$& $-n$ \\
		\hline $(0  \cdots 0)_\mathrm{s}$ & $\Omega$& $-2n$ \\ \hline 
\end{tabular}  
\caption{Highest-weight vectors for the irreducible submodules in the decomposition of  (\ref{eq:sundec}). The eigenvalues of the Casimir element have been computed previously~\cite{MacfarlanePfeiffer00}. Furthermore, $v \wedge w = v \otimes w - w \otimes v$ and $\mathrm{Sym}(v \otimes w) = v \otimes w + w \otimes v$.} 
\label{tab:hwv2}
\end{table}

We note that the highest-weight vectors for the two adjoint representations in \eqref{eq:sundec} are normalised such that they have the same norm with respect to the inner product on $\mathrm{su(\emph{n})}\otimes\mathrm{su(\emph{n})}$ given for pure tensor states by $(v_1\otimes v_2,w_1\otimes w_2)=(v_1,w_1)\,(v_2,w_2)$. Writing the two highest-weight vectors as $v_\mathrm{s} = n \sum_{a,b,c} w^a d^{abc} I^b \otimes I^c$ and $v_\mathrm{a} = \sqrt{n^2-4} \sum_{a,b,c} w^a f^{abc} I^b \otimes I^c$, where $w^a$ is the unique vector such that $e^{1n} = \sum_a w^a I^a$, the norms can be computed easily using~\cite{Macfarlane-68} $\sum_{a,b}d^{abc} d^{abd} = (2n^2-8)\delta^{cd}/n$ and $\sum_{a,b}f^{abc} f^{abd} = -2n\delta^{cd}$.

\section{Construction of the R-matrix in the adjoint representation}\label{sec:Rmat}
In this section we derive the R-matrix in the adjoint representation from the corresponding intertwining operator. A similar construction was performed in Ref.~\cite{ChariPressley91} for the representation $g\otimes\mathbb{C}$ if $g\neq$ su(\emph{n}). 

Consider the two representations $V_\lambda$ and $V_\mu$ of Y(su(\emph{n})) obtained by pulling back the adjoint representation $V=(10\ldots 01)$ of su(\emph{n}) using the evaluation homomorphism \eqref{eq:evahom} with the spectral parameters $\lambda$ and $\mu$ respectively. The action of Y(su(\emph{n})) on the tensor product $V_\mu\otimes V_\lambda$ is defined by the co-multiplication \eqref{eq:comultiplication}. As was proven by Chari and Pressley~\cite{ChariPressley91}, there exists a unique intertwining operator $I(\lambda-\mu): V_\mu\otimes V_\lambda\to V_\lambda\otimes V_\mu$ compatible with this representation, ie, $I$ satisfies 
\begin{equation}
\fl\quad
I(\lambda-\mu)\,(\ad_\mu\otimes\ad_\lambda)\Delta(y)=(\ad_\lambda\otimes\ad_\mu)\Delta(y)\,I(\lambda-\mu),\quad y\in \mathrm{Y(su(\emph{n}))},
\label{eq:intertwining}
\end{equation}
that preserves the tensor product of the highest-weight states. For later use we give the explicit action of the $J(x)$ on $V_\mu\otimes V_\lambda$,
\begin{eqnarray}
\fl\quad
J(x)_{\mu,\lambda}=(\ad_\mu\otimes\ad_\lambda)\Delta(J(x))=\left(\mu x+\frac{1}{4}\sum_{a,b}\mathrm{tr}\Bigl(x\{I^a,I^b\}\Bigr)I^a I^b\right)\otimes 1\nonumber\\*
\qquad\qquad\qquad+1\otimes \left(\lambda x+\frac{1}{4}\sum_{a,b}\mathrm{tr}\Bigl(x\{I^a,I^b\}\Bigr)I^a I^b\right)+\frac{1}{2}\Bigl[x\otimes 1,\Omega\Bigr].
\label{eq:actionJ}
\end{eqnarray}
We stress that the action of products like $I^aI^b$ has to be understood in the adjoint representation, ie, $I^aI^b\equiv\ad(I^a)\,\ad(I^b)=[I^a,[I^b,\,.\,]]$. 

We first note that since $I$ is a function of $\lambda-\mu$, we can set $\mu=0$. Furthermore, the intertwining operator is invariant under the action of the Lie algebra su(\emph{n}), thus the decomposition \eqref{eq:sundec} together with Schur's lemma implies that it can be written as 
\begin{eqnarray}
I(\lambda) &=& P_{(20\ldots 02)} + f_1(\lambda)P_{(20\ldots 010)} + f_2(\lambda)P_{(010\ldots 02)}+ f_3(\lambda)P_{(010\ldots 010)} \nonumber \\ 
&&+ f_4(\lambda)P_{(0\ldots 0)} + M(\lambda),
\label{eq:ISchur}
\end{eqnarray}
where $P_\Lambda$ is the projection onto the submodule $\Lambda$, the $f_i(\lambda)$ are rational functions, and $M(\lambda)$ is an operator acting on the two copies of the adjoint representation; when considering its action on the two highest-weight states (see Table~\ref{tab:hwv2}) it can be written as a $(2 \times 2)$-matrix. Note that the requirement to preserve the tensor product of the highest-weight states fixes the prefactor of $P_{(20\ldots 02)}$ to one.

The functions $f_i(\lambda)$ and the entries of $M(\lambda)$ can be obtained recursively from the general relation \eqref{eq:intertwining}. Starting with the highest-weight state $v=v_{(20\ldots 010)}$ of the representation $(20\ldots 010)_\mathrm{a}$ in \eqref{eq:sundec}, the requirement \eqref{eq:intertwining} for the element $y=J(e_{(n-1)n})$ yields 
\begin{equation}
I(\lambda)\,J(e_{(n-1)n})_{0,\lambda}\,v_{(20\ldots 010)}=J(e_{(n-1)n})_{\lambda,0}\,I(\lambda)\,v_{(20\ldots 010)},
\end{equation}
where the action of $J(e_{(n-1)n})_{\mu,\lambda}$ is given by \eqref{eq:actionJ}. This can be rewritten as 
\begin{eqnarray}
\fl\quad I(\lambda)\left[\lambda\,\Bigl(1\otimes e_{(n-1)n}\Bigr)+\frac{1}{2}\Bigl(e_{(n-1)n}\otimes 1\Bigr)\,\Omega-\frac{1}{2}\Omega\,\Bigl(e_{(n-1)n}\otimes 1\Bigr)\right]v_{(20\ldots 010)}\nonumber\\*
=f_1(\lambda)\left[\left(\lambda-\frac{1}{2}\Omega\right)\,\Bigl(e_{(n-1)n}\otimes 1\Bigr)+\frac{1}{2}\Bigl(e_{(n-1)n}\otimes 1\Bigr)\,\Omega\right]v_{(20\ldots 010)},
\end{eqnarray}
where we used the identity
\begin{equation}
\fl\quad\sum_{a,b}\Bigl[\mathrm{tr}\Bigl(e_{(n-1)n}\{I^a,I^b\}\Bigr)\,I^aI^b\otimes 1+1\otimes \mathrm{tr}\Bigl(e_{(n-1)n}\{I^a,I^b\}\Bigr)\,I^aI^b\Bigr]v_{(20\ldots 010)}=0.
\label{eq:identity1}
\end{equation}
We have checked \eqref{eq:identity1} by explicit numerical evaluation for $n\le 7$. Now using
\begin{equation}
\Bigl(1\otimes e_{(n-1)n}\Bigr)v_{(20\ldots 010)} = -\Bigl(e_{(n-1)n} \otimes 1\Bigr)v_{(20\ldots 010)} = v_{(20\ldots 02)}
\end{equation}
as well as $\Omega\,v_{(20\ldots 010)}=0$ and $\Omega\,v_{(20\ldots 02)}=2v_{(20\ldots 02)}$ we obtain
\begin{equation}
(1+\lambda)v_{(20\ldots 02)}=(1-\lambda)\,f_1(\lambda)\,v_{(20\ldots 02)}\quad\Rightarrow\quad f_1(\lambda)=\frac{1+\lambda}{1-\lambda}.
\end{equation}
The same calculation can be performed for $v=v_{(010\ldots 02)}$ and $y=J(e_{12})$. As the identity \eqref{eq:identity1} with $e_{(n-1)n}$ and $v_{(20\ldots 010)}$ replaced by $e_{12}$ and $v_{(010\ldots 02)}$, respectively, still applies (also explicitly checked for $n\le 7$) we find $f_2(\lambda)=f_1(\lambda)$. Similarly, for $v=v_{(010\ldots 010)}$ and $y=J(e_{12})$ the function $f_3(\lambda)$ can be related to $f_1(\lambda)$ with the result $f_3(\lambda)=(1+\lambda)^2/(1-\lambda)^2$. The last function, $f_4(\lambda)$, can be obtained by acting with $J(e_{1n})$ twice on the highest-weight state $v_{(0\ldots 0)}=\Omega$. Using the identity
\begin{equation}
J(e_{1n})_{\mu,\lambda}\,J(e_{1n})_{\mu,\lambda}\,v_{(0\ldots 0)}=-2(\mu-\lambda-1)(\mu-\lambda-n)\,v_{(20\ldots 02)},
\end{equation}
again explicitly checked for $n\le 7$, we obtain $f_4(\lambda)=(1+\lambda)(n+\lambda)/(1-\lambda)/(n-\lambda)$. 

In order to determine the intertwining operator on the submodule $(10\ldots 01)_\mathrm{s}\oplus (10\ldots 01)_\mathrm{a}$, we consider its action on the corresponding highest-weight states $v_{(10\ldots 01)_\mathrm{s}}$ and $v_{(10\ldots 01)_\mathrm{a}}$ respectively. The action on these is encoded in the $(2\times 2)$-matrix $M(\lambda)$, whose entries can be calculated from \eqref{eq:intertwining} using the set of identities
\begin{eqnarray}
\fl\quad J(e_{1n})_{\mu,\lambda}\,v_{(10\ldots 01)_\mathrm{s}}=n(n- 2)\,v_{(20\ldots 02)},\label{eq:ex6}\\
\fl\quad J(e_{1n})_{\mu,\lambda}\,v_{(10\ldots 01)_\mathrm{a}}=\sqrt{n^2-4}\,\Bigl(n + 2 -2 \mu + 2 \lambda\Bigr)\,v_{(20\ldots 02)},\label{eq:ex7}\\
\fl\quad J\left(e_{(n-1)n}\right)_{\mu,\lambda}\,J\left(e_{1(n-1)}\right)_{\mu,\lambda}\,v_{(10\ldots 01)_\mathrm{s}}=n\left[(\lambda - \mu)^2 + 2\lambda - n\mu + \frac{n^2}{4}\right]v_{(20\ldots 02)}, \label{eq:ex8}\\
\fl\quad J\left(e_{(n-1)n}\right)_{\mu,\lambda}\,J\left(e_{1(n-1)}\right)_{\mu,\lambda}\,v_{(10\ldots 01)_\mathrm{a}}\nonumber\\
\fl\qquad =\frac{\sqrt{n^2-4}}{4}\Bigl[n(n+4) - (6n+4)\mu + 4\mu^2 + (2n-4)\lambda -4\lambda^2\Bigr]v_{(20\ldots 02)}.\label{eq:ex9}
\end{eqnarray}
As before, we have checked \eqref{eq:ex6}--\eqref{eq:ex9} by explicit numerical evaluation for $n\le 7$. The matrix $M(\lambda)$ is now obtained by straightforward calculation with the result 
\begin{equation}
\fl\quad
M(\lambda) = \frac{1}{2(n-\lambda)(1-\lambda)^2}\left(\begin{array}{cc} 2n + (n^2+2)\lambda-2\lambda^3 & n\sqrt{n^2 - 4}\lambda \\   -n \sqrt{n^2-4}\lambda & 2n - (n^2+2)\lambda + 2\lambda^3 \end{array}\right),
\label{eq:M}
\end{equation}
where we recall that $M(\lambda)$ has to be understood with respect to the ordered basis of highest-weight vectors $\{v_{(10\ldots 01)_\mathrm{s}},v_{(10\ldots 01)_\mathrm{a}}\}$. One easily checks $I(\lambda)\,I(-\lambda)=1$. This completes our derivation of the intertwining operator \eqref{eq:ISchur}.

The R-matrix is now obtained by the composition of the intertwining and permutation operators. Using the symmetry or antisymmetry of the submodules as indicated in \eqref{eq:sundec}, we arrive at the R-matrix in the adjoint representation of su(\emph{n}), which constitutes our main result (we recall that $n\ge 4$)
\begin{eqnarray}
\fl \quad R(\lambda) = I(\lambda)\,\sigma=P_{(20\ldots 02)} - \frac{1+\lambda}{1-\lambda}\Bigl(P_{(20\ldots 010)} + P_{(010\ldots 02)}\Bigr)+ \left(\frac{1+\lambda}{1-\lambda}\right)^2\,P_{(010\ldots 010)} \nonumber \\ 
\qquad+\frac{1+\lambda}{1-\lambda}\frac{n+\lambda}{n-\lambda}\,P_{(0\ldots 0)} + N(\lambda).
\label{eq:rmat}
\end{eqnarray}
Here the operator $N(\lambda)$, considered as acting on the highest-weight states of $(10\ldots 01)_\mathrm{s}$ and $(10\ldots 01)_\mathrm{a}$, is given by 
\begin{equation}
\fl\quad
N(\lambda) = \frac{1}{2(n-\lambda)(1-\lambda)^2}\left(\begin{array}{cc} 2n + (n^2+2)\lambda-2\lambda^3 & -n\sqrt{n^2 - 4}\,\lambda \\   -n \sqrt{n^2-4}\,\lambda & -2n + (n^2+2)\lambda - 2\lambda^3 \end{array}\right).
\label{eq:N}
\end{equation}
The R-matrix satisfies $R(0) = \sigma$ as well as 
\begin{equation}
R(\lambda)=\left(1+\frac{2}{\lambda}\right)\,1\otimes 1-\frac{1}{\lambda}(\ad\otimes\ad)\,\Omega+\mathcal{O}(\lambda^{-2}),
\end{equation}
furthermore $\tilde{R}(\lambda)=\sigma\,I(\lambda)$ provides a second, independent solution. We note that the construction followed above cannot be generalised to the adjoint representation of other Lie algebras $g\neq\mathrm{su(\emph{n})}$. Instead one has to consider the representation $g\otimes\mathbb{C}$, as was done in Ref.~\cite{ChariPressley91}. The obtained results are similar to \eqref{eq:rmat}. 

\section{Integrable SU(\emph{n}) spin chains}\label{sec:hamiltonian}
The quantum inverse scattering method~\cite{KorepinBogoliubovIzergin93,SamajBajnok13} allows the construction of integrable models from a given solution of the Yang--Baxter equation. In the case at hand we obtain a spin chain with adjoint SU(\emph{n}) symmetry. Specifically consider an $N$-site chain with periodic boundary conditions. Each lattice side carries the $(n^2-1)$-dimensional, adjoint representation of su($n$). Since the R-matrix \eqref{eq:rmat} satisfies $R(0) = \sigma$, the integrable Hamiltonian is given by~\cite{SamajBajnok13}
\begin{equation}
H = \sum\limits_{i=1}^N h_{i,i+1},\quad h=\frac{\partial}{\partial\lambda}R(\lambda)\Big|_{\lambda=0}\,\sigma,
\label{eq:ham}
\end{equation}
where $h_{i,i+1}$ denotes the local Hamiltonian acting non-trivially only on the two neighbouring sites $i$ and $i+1$. We find
\begin{equation}
h_{i,i+1}=2\Bigl(P_{(20\ldots 010)} + P_{(010\ldots 02)}\Bigr)+4\,P_{(010\ldots 010)}+\frac{2+2n}{n}\,P_{(0\ldots 0)}+O,
\label{eq:hlocal}
\end{equation}
where $O$, written as a matrix acting on the two copies of the adjoint representation, is given by
\begin{equation}
O = \left(\begin{array}{cc}\frac{(2+n)^2}{2n} & \frac{1}{2}\sqrt{n^2-4} \\-\frac{1}{2}\sqrt{n^2-4} & 2 - \frac{n}{2}  \end{array}\right).
\end{equation}
We recall that the matrix is given in the basis of Table~\ref{tab:hwv2}. 

It is well known that the Hamiltonians of the SU(2)-invariant integrable models~\cite{Babujian83} constructed using the quantum inverse scattering method can be expressed in terms of a polynomial in the Casimir element. For the adjoint representation one obtains the integrable Takhtajan--Babujian model~\cite{Takhtajan82,Babujian83}. In contrast, because of the off-diagonal terms, it is not possible to rewrite \eqref{eq:hlocal} in terms of a polynomial in the Casimir element. 

Nevertheless, we can express the local Hamiltonian in terms of SU(\emph{n}) 'spin' operators acting on the lattice sites. These spin operators for the adjoint representation can be defined as follows: We fix an orthonormal basis $I^a$ with respect to the trace form. Then the spin operators act in the adjoint representation by a matrix $S^a=\ad(I^a)$ with the structure constants as entries, ie, $(S^a)_{bc}=f^{abc}$. Since in the basis $\{I^a\}$ the structure constants are completely antisymmetric and purely imaginary, the spin operators $S^a$ are Hermitian. Furthermore, we define the following Hermitian operators~\cite{RoyQuella15}
\begin{eqnarray}
Q = \sum_a S^a_1 S^a_{2},\quad C_\mathrm{A} = \sum_{a,b,c} d^{abc} \Bigl(S^a_1 S^b_1 S^c_{2} - S^a_1 S^b_2 S^c_{2} \Bigr),\nonumber\\
K = \sum_{a,b,c,d,e,f} d^{abc} d^{def} S^a_1 S^d_1 S^e_1 S^f_{2} S^b_{2} S^c_{2},
\label{eq:QCKdef}
\end{eqnarray}
where the spin operators act on two neighbouring sites indicated by the subindex 1,2. Applying certain relations~\cite{RoyQuella15} between these operators, the local Hamiltonian can be written as~\cite{footnote1} (we recall that $n>3$)
\begin{eqnarray} 
\label{eq:localhamsunexpr}
h_{i,i+1} = Q +\frac{1}{n}\frac{2-n^2}{6+n^2}Q^2 - \frac{3}{6+n^2}Q^3 -\frac{1}{n}\frac{1}{6+n^2}Q^4\nonumber\\*
\qquad\qquad-\frac{1}{n^3}\frac{2 + n^2}{6+n^2}K + \frac{1}{4n^3}\frac{1}{6+n^2} \Bigl[K,C_\mathrm{A}\Bigr],\end{eqnarray}
where $C$, $C_\mathrm{A}$ and $K$ are understood to act on sites $i$ and $i+1$. We note that in the derivation of \eqref{eq:localhamsunexpr} we rescaled \eqref{eq:hlocal} by the factor $(6+n^2)/8$ and dropped an additive constant. From this result we immediately deduce that the local Hamiltonian is not Hermitian because of the commutator $[K,C_\mathrm{A}]$, and thus that also the total Hamiltonian \eqref{eq:ham} will be non-Hermitian. We note that while the eigenvalues of the two-site system $h_{i,i+1}$ are real, for the full Hamiltonian this is no longer the case. Hence the physical interpretation of the system \eqref{eq:ham} is unclear. 

\section{Special case: SU(3) model}\label{sec:SU3}
Finally we consider the special case of SU(3) which has been investigated previously~\cite{AlishauskasKulish86} without using the Yangian. The decomposition of the tensor product \eqref{eq:sundec} does not contain the representation $(010\ldots 010)$. Denoting the representations by their dimensions, eg, $(30)=\mathbf{10}$, we thus have instead of \eqref{eq:sundec}
\begin{equation} \label{eq:8dec}
  \mathbf{8} \otimes \mathbf{8}  = \mathbf{27} \oplus \mathbf{10} \oplus \mathbf{\bar{10}} \oplus \mathbf{8}_\mathrm{s} \oplus \mathbf{8}_\mathrm{a} \oplus \mathbf{1}.
\end{equation}
Again the subindices of the adjoint representations indicate its symmetry/antisymmetry under exchange of the factors, while the bar denotes conjugation. The highest-weight vectors for these modules are still given in Table~\ref{tab:hwv2}, while the R-matrix reads
\begin{equation}
R(\lambda)=P_\mathbf{27} - \frac{1+\lambda}{1-\lambda}\Bigl(P_\mathbf{10} + P_\mathbf{\bar{10}}\Bigr)+\frac{1+\lambda}{1-\lambda}\frac{3+\lambda}{3-\lambda}\,P_\mathbf{1}+N(\lambda)
\label{eq:rmat3}
\end{equation}
with $N(\lambda)$ given by setting $n=3$ in \eqref{eq:N}. The matrix $N(\lambda)$ can be diagonalised with the eigenvalues given by $(11\lambda-2\lambda^2\pm 3\sqrt{4+5\lambda^2})/2/(1-\lambda)^2/(3-\lambda)$ in agreement with Ref.~\cite{AlishauskasKulish86}. Furthermore, we have explicitly checked that the R-matrix \eqref{eq:rmat3} satisfies the Yang--Baxter equation. The local Hamiltonian becomes
\begin{equation}
h_{i,i+1} = 2 \Bigl(P_\mathbf{10} + P_{\bar{\mathbf{10}}}\Bigr)  + \frac{25}{6}\,P_{\mathbf{8}_\mathrm{s}} +   \frac{1}{2}\,P_{\mathbf{8}_\mathrm{a}} +\frac{\sqrt{5}}{2} \Bigl(O_\mathrm{sa} -O_\mathrm{as}\Bigr) + \frac{8}{3}\,P_\mathbf{1},
\label{eq:hsu3}
\end{equation}
where $O_\mathrm{as}$ is mapping $\mathbf{8}_\mathrm{s}$ onto $\mathbf{8}_\mathrm{a}$ and vice versa. When considered as acting on the highest-weight states we have explicitly $O_\mathrm{as}=\sigma^-$ and $O_\mathrm{sa}=\sigma^+$ with the Pauli matrices $\sigma^\pm=(\sigma^\mathrm{x}\pm\mathrm{i}\sigma^\mathrm{y})/2$. Introducing Hermitian spin operators $S^a$ via $S^a=\ad(I^a)$ with $\{I^a\}$ an orthonormal  basis of su(3), eg, $I^a=\lambda^a/\sqrt{2}$ with $\lambda^a$ being the Gell-Mann matrices, the local Hamiltonian can be written as 
\begin{equation}
h_{i,i+1} = Q - \frac{7}{9}Q^2-\frac{2}{9}Q^3-\frac{11}{81}K+\frac{1}{324} \Bigl[K,C_\mathrm{A}\Bigr],
\end{equation}
where the operators $Q$, $K$ and $C_\mathrm{A}$ were defined in \eqref{eq:QCKdef}, and we have rescaled \eqref{eq:hsu3} by 8/3. We note that due to the last term the full Hamiltonian is not Hermitian.

\section{Conclusion}\label{sec:discussion}
We constructed the intertwining operator between two representations of the Yangian of su(\emph{n}) obtained by its  pull-back of the adjoint representation. From this we obtained the rational R-matrix with adjoint SU(\emph{n}) symmetry, given in \eqref{eq:rmat}, as well as the corresponding integrable Hamiltonian. From the physics point of view the main problem with the latter is its lack of hermiticity, which renders its physical interpretation unclear. Still we hope that the derived R-matrix will be useful in other contexts like field theories with factorised scattering or quantum information theory.

\ack

We would like to thank Paul Fendley, Frank G\"ohmann, Tatjana Pu\v{s}karov, Thomas Quella and Nicolai Reshetikhin for valuable comments and discussions. This work is part of the D-ITP consortium, a program of the Netherlands Organisation for Scientific Research (NWO) that is funded by the Dutch Ministry of Education, Culture and Science (OCW). 

\section*{References} 


\begin{thebibliography}{10}
\bibitem{footnote1}
The sign in front of the commutator was not fixed in Ref.~\cite{RoyQuella15}, but for $n\le 7$ we fixed it by explicit calculation.
\end{thebibliography}

\end{document}